\documentclass{PoS}

\title{Accurate calculations of the WIMP halo around the Sun and prospects for gamma ray detection}

\ShortTitle{The WIMP halo around the Sun}

\author{\speaker{Sofia Sivertsson}\\ 
        Department of Theoretical Physics, Royal Institute of Technology (KTH)\\
	AlbaNova University Center, 106 91 Stockholm, Sweden\\
        E-mail: \email{sofiasi@kth.se}}

\author{Joakim Edsj\" o\\
        Department of Physics, Stockholm University\\
	AlbaNova University Center, 106 91 Stockholm, Sweden\\
        E-mail: \email{edsjo@physto.se}}

\abstract{Weakly interacting massive particles (WIMPs) can be captured by heavenly objects, like the Sun. Under the process of being captured by the Sun, they will build up a population of WIMPs around it, that will eventually sink to the core of the Sun. It has been argued with simpler estimates before that this halo of WIMPs around the Sun could be a strong enough gamma ray source to be a detectable signature for WIMP dark matter. We here revisit the problem using detailed Monte Carlo simulations and detailed composition and structure information about the Sun to estimate the size of the gamma ray flux. Compared to earlier estimates, we find that the gamma ray flux from WIMP annihilations in the Sun halo would be negligible and no current or planned detectors would even be able to detect this flux.\ }

\FullConference{Identification of dark matter 2008\\
		 August 18-22, 2008\\
		 Stockholm, Sweden}

\usepackage{epsfig}
\begin{document}
\section{Introduction}
\subsection{Previous work}

Earlier estimates of this signal made by Strausz in 1998 showed that this gamma ray signal would be detectable or, if no such signal is seen, constrain parameters in the WIMP model \cite{Strausz}. Hooper revisited the problem in 2001 and found the gamma ray signal to be many orders of magnitude lower, too low to be detectable with realistic telescope areas \cite{Hooper}. In 2003 Fleysher performed a slightly more detailed calculation and found even higher rates than Strausz \cite{Fleysher}. Also, the Milagro detector has made a search for this gamma ray signal in 2004 without any success \cite{milagro}. 

The aim of this work is to make a much more detailed analysis of the WIMP density around the Sun in order to achieve a more accurate estimate of the gamma ray signal at earth.

\subsection{The WIMP capture process}

For the Milky Way having a smooth WIMP halo, some fraction of the WIMPs passing through the Sun will scatter off nuclei in the Sun. If enough energy is lost in the scatter the WIMP becomes unable to escape the Sun's gravitational well and ends up in some orbit around the centre of the Sun. These bound WIMPs will eventually scatter again and as this energy loss continues the WIMP will eventually end up in an orbit which is completely hidden inside the Sun.

The WIMPs are typically much heavier than the nuclei they scatter off, giving small energy losses in each scatter, and hence often require a large number of scatters before the WIMP is completely hidden inside the Sun. Also, the smallness of the scatter cross section typically makes a bound WIMP survive many passages in the sun before it scatters again. Due to this the capture process typically takes quite long time and all these intermediate bound WIMPs orbiting the Sun will give rise to an overdensity of dark matter, i.e. a WIMP halo around the Sun.

WIMP annihilations within this halo give rise to high energy gamma rays benefiting from a very low background since the Sun does not emit photons with such high energies. Also, the Sun is opaque to gamma rays and hence shields against the diffuse gamma ray background.

\section{Calculation of the density of the Sun's WIMP halo.}
The WIMP density around the Sun is calculated using a Monte Carlo simulating the capture process of a large number of WIMPs, in combination with analytic calculations. The Monte Carlo takes the full properties of the WIMP orbits into account and is constructed as described in the following text.

\subsection{The WIMPs' first scatter in the Sun}
In the relevant work by Gould \cite{Gould} the total number of WIMPs scattering at a given radius in the Sun per unit time and velocity is determined. The WIMPs in the Milky Way halo are assumed to have Maxwell-Boltzmann distributed velocities with the velocity dispersion $\bar v=270$ km/s. For a smooth halo the WIMP density in our region of the Milky Way is 0.3 GeV cm$^{-3}$.

Using the work by Gould and the kinematic laws of elastic scatter, taking into account cross section dependence on energy loss in the scatter, one can derive the energy distribution of the scattered WIMPs after their first scatter in the Sun, given as a function of where in the Sun the scatter took place. Energy here refers to the WIMPs total energy (kinetic plus potential) and the scattered WIMP is hence captured if scattered to an energy less than zero. Integrating this energy distribution after the WIMPs' first scatter over energies less than zero yields the total capture rate of WIMPs in the Sun. The total capture rate has also been calculated in DarkSUSY \cite{darksusy} for various parameter settings and is in excellent agreement with our result.

\subsubsection{The captured WIMPs after their first scatter}
To continue the calculation from this point it is required to start looking at individual WIMPs, e.g. to write a Monte Carlo. In the Monte Carlo the starting point is to pick an energy and a scatter radius according to the distribution discussed above. Due to the spherical symmetry of the Sun the velocities of the scattered WIMPs are assumed to be isotropic in directions, also the smallness of the scatter probability makes it effectively equally probable for a WIMP to scatter on its way out as on its way in through the Sun. Randomly picking the direction which the WIMP scatters to then gives, together with the energy and place of scatter, the angular momentum of the WIMP's new orbit.

The energy and angular momentum, together with the gravitational potential, fully specifies the orbit of the bound WIMP. Outside the Sun the orbits are truly elliptical and the density contribution from one lap in the given orbit is then determined by how much time a particle in an elliptical orbit spends at different distances from one of the foci.

To determine the WIMP's total contribution to the dark matter density before it scatters again the scatter probability as the WIMP traverses the Sun is required. In calculating the scatter probability for a solar passage we take into account that the orbits are not truly elliptical inside the Sun. The true orbit are slightly more stable since they do not come as close to the centre of the Sun as a perfect ellipse would. Also, it is taken into account that for scatter off heavier nuclei the scatter cross section depends on the energy loss in the scatter, which depends on the WIMP velocity and hence the radius of the scatter.

\subsection{Subsequent scatters}
The radius of scatter of the bound WIMP is picked according to the distribution required for the total scatter probability and after this the type of element which the WIMP scatters off is determined. The WIMP's energy loss in the scatter is then determined and the new orbit is specified. The density contribution from this new orbit is added to the total WIMP density and the whole process is repeated as the WIMP is scattered again and again until complete solar entrapment. The capture process is simulated for a large number of WIMPs and then normalized with the total capture rate.

\section{Results}

\begin{figure}
\centerline{\epsfig{file=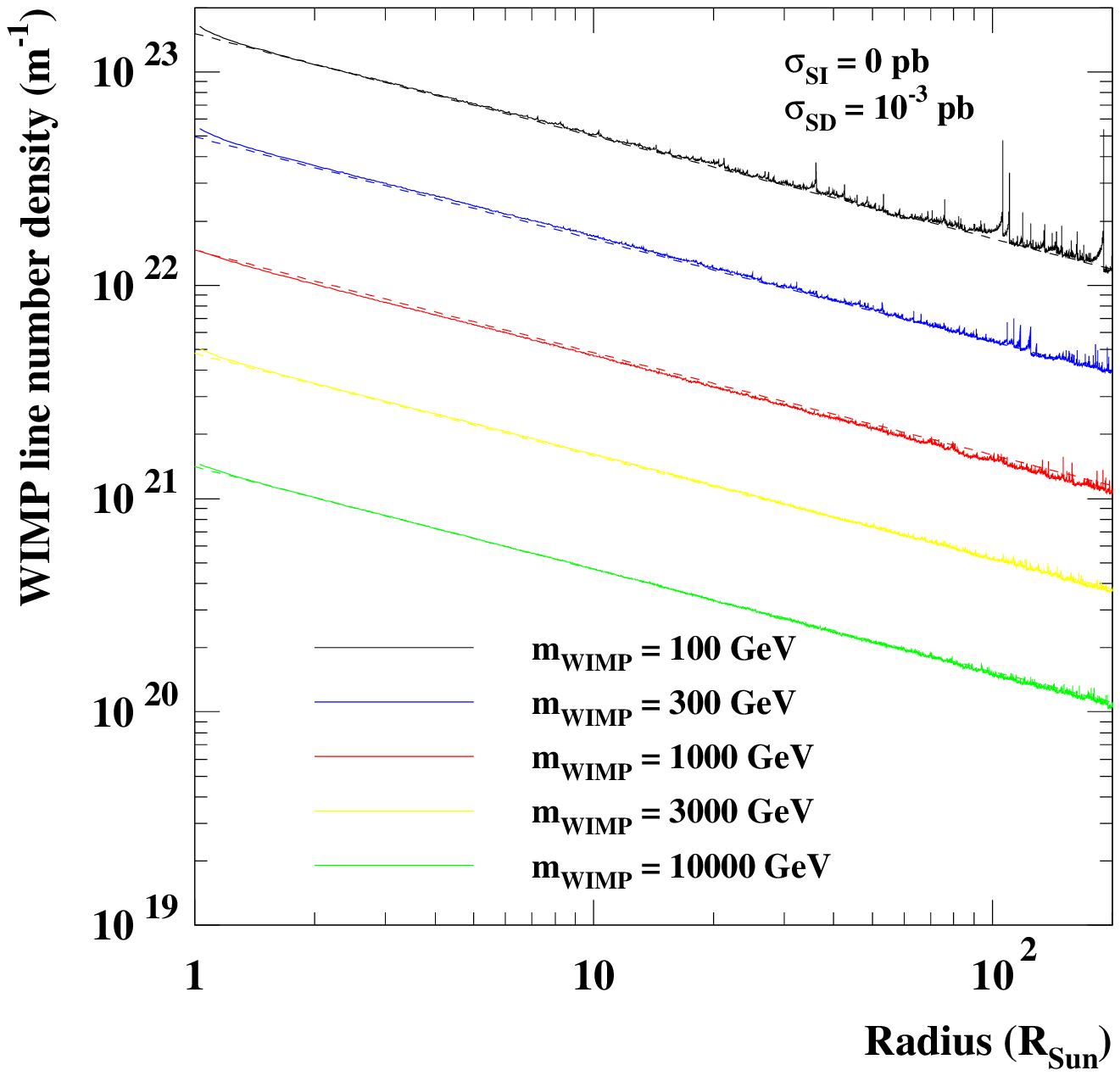,width=0.49\textwidth}\epsfig{file=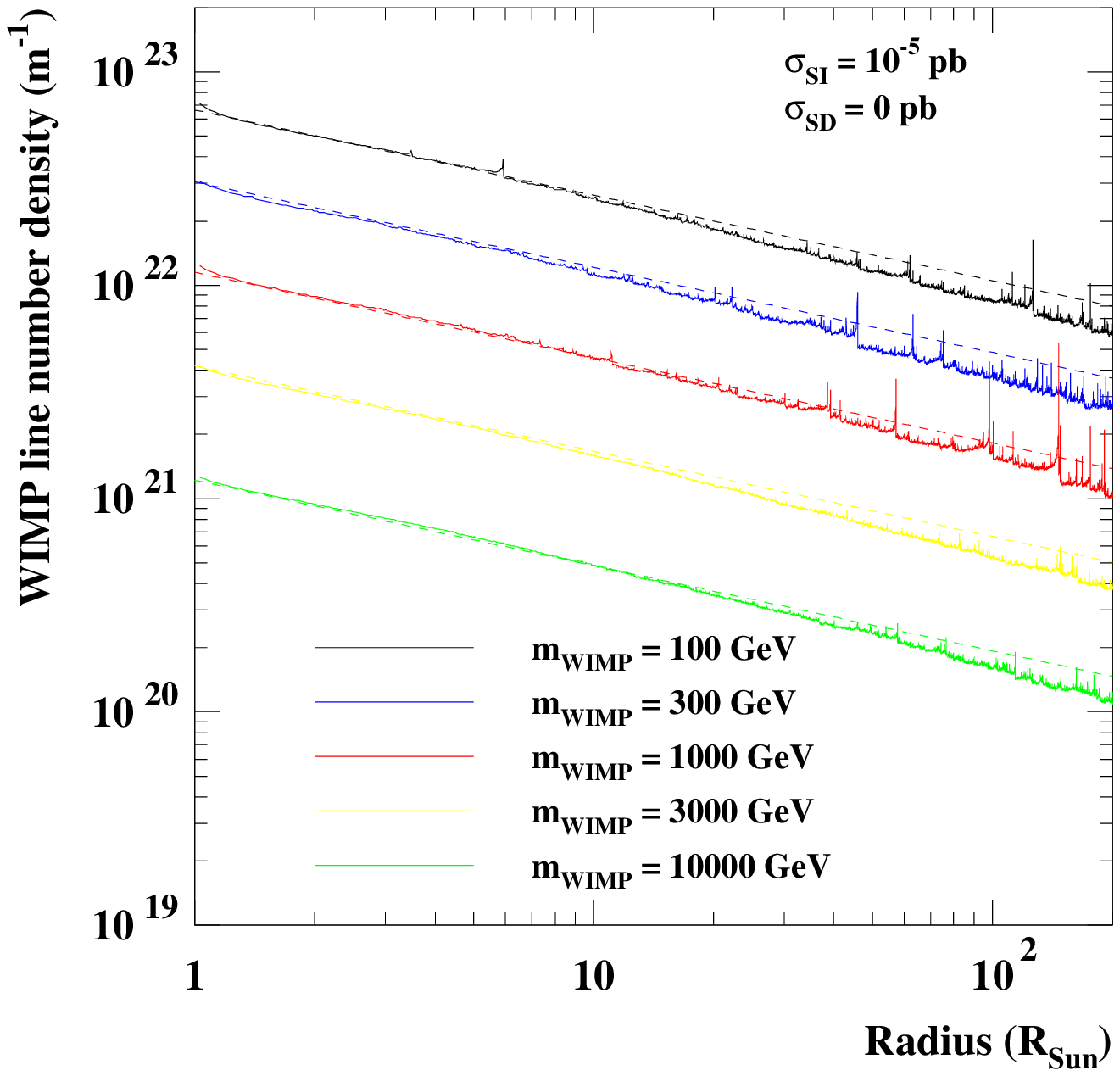,width=0.49\textwidth}}
\caption{The line number density of WIMPs around the Sun with spin dependent cross section being dominant (only scatters off hydrogen) and the spin independent cross section being dominant, respectively. In the left graph $\sigma_{SD}=10^{-3}$ pb and $\sigma_{SI}=0$ pb, and $\sigma_{SD}=0$ pb and $\sigma_{SI}=10^{-5}$ pb for the right. The dashed curves are the fits described by Eq~(\protect\ref{eq:ns}).}
\label{fig:linedens}
\end{figure}

The Monte Carlo simulations of around one million captured WIMPs yields the WIMP number density around the Sun as shown in Figure~\ref{fig:linedens}. Which elements in the Sun the WIMPs scatter off depends on the cross section configuration, here both the cases of spin dependent, $\sigma_{SD}$, and spin independent, $\sigma_{SI}$, cross sections being dominant have been studied.

As shown in the graphs in Figure~\ref{fig:linedens} the WIMP number density can be well approximated by
\begin{eqnarray}
n^{SD} & = & 10^{25.21 - 1.015x} \left(\frac{r}{r_\odot}\right)^{-0.48} \mbox{ m$^{-1}$}\nonumber \\
n^{SI} & = & 10^{23.71 - 0.2332x -0.1056x^2} \left(\frac{r}{r_\odot}\right)^{-0.40} \mbox{ m$^{-1}$} \label{eq:ns}\\
\mbox{with } x & = & \log_{10} \left( \frac{m_{WIMP}}{\mbox{1 GeV}} \right)\nonumber .
\end{eqnarray}

For the simple choice of WIMP annihilation cross section of $\sigma v=10^{-32}$ m$^3$s$^{-1}$ and number of gamma rays produced per WIMP annihilation to be $N_{\gamma}=20$, independent of WIMP mass. The gamma ray flux at Earth is then, for some WIMP masses, given in Table \ref{table:flux}.

\begin{table}[ht]
\centering 
\begin{tabular}{|c|ccc|r|}
	\hline
$ $ &  $m_{WIMP}=100$ GeV    &   $m_{WIMP}=1$ TeV  & $m_{WIMP}=10$ TeV \\
	\hline
$\sigma_{SD}=10^{-3}$ pb, $\sigma_{SI}=0$  & $4.0\cdot 10^{-19}$  & $3.7\cdot 10^{-21}$ & $3.5\cdot 10^{-23}$ \\
$\sigma_{SD}=0$, $\sigma_{SI}=10^{-5}$ pb & $8.4\cdot 10^{-20}$  & $2.5\cdot 10^{-21}$ & $2.9\cdot 10^{-23}$ \\
	\hline
\end{tabular}
\caption{The total flux (photons per $\mbox{m}^2$ per second) at Earth of gamma rays from the Sun's WIMP halo.}
\label{table:flux}
\end{table}

The WIMP annihilation rate is much lower than the capture rate and hence has no significant impact on the WIMP density.
As one might have noticed the density of the WIMP halo is independent of the magnitude of the scatter cross section. A lower scatter cross section makes the Sun less capable of capturing WIMPs but on the other hand the capture process takes longer since it makes the WIMP orbits more stable. These two effects exactly cancel except for very small scatter cross sections where the finite age of the Sun starts to become important. For example looking at $100$ GeV WIMPs with $\sigma_{SD}=10^{-3}$ pb and $\sigma_{SI}=0$, a reduction of the scatter cross section by five orders of magnitude lowers the WIMP density in the region close to the Sun by roughly 20\% due to the finite age of the Sun.

\section{Discussion}

Part of the reduction of the gamma ray signal for the heavy WIMPs in Table~\ref{table:flux} is an artifact of the assumption of $N_\gamma$ being independent of the WIMP mass. Heavy WIMPs are more difficult for the Sun to capture but they also require more scatters to be trapped inside the Sun. Heavy WIMPs with low scatter cross sections are hence potentially more sensitive to the finite age of the Sun.

In this calculation the effects of the planets of our solar system have not been taken into account. The planets can accelerate and throw orbiting WIMPs out of the solar system. WIMPs orbiting the Sun can also be disturbed in such a way that their orbits no longer intersect the Sun, making them more stable but eventually also thrown out. Both these effects will, if they are of any significance, reduce the number of WIMPs reaching the low energy orbits. The gamma ray production is dominated by the region close to the Sun, where the low energy orbits are the most important, since orbits of higher energies spend very little time close to the Sun.

The gamma ray signal, as seen in Table~\ref{table:flux}, is truly very low. The calculated flux corresponds to, at best, less than one gamma photon per 100 km$^2$ per century, which is hardly ever detectable. For the heavy WIMP scenario the high energy gamma rays could in theory benefit from the large area of air Cherenkow telescopes, however these cannot look towards the Sun since the light from the Sun would spoil the Cherenkow light signal. Other telescopes, such as Fermi and Milagro, have way to small surface areas. The unfortunate conclusion is hence that this signal will not be observable now nor in the foreseeable future.

The low signal in this work contradicts the result by Strausz \cite{Strausz}. The analysis by Strauss is not as detailed as the work here, he assumes one-dimensional orbits and treats the WIMP orbit properties in an average sense. However, those simplifying assumptions should not in themselves cause such a large disagreement and the origin of this disagreement is not clear. Our conclusion agrees with the more estimative calculation by Hooper \cite{Hooper} even though our concluded gamma ray flux is slightly lower than his.

\end{document}